\documentclass{aastex631}

\graphicspath{{./}{figures/}}

\usepackage{amssymb}
\usepackage{amsmath}
\usepackage{cases}

\begin{document}

\title{When hot meets cold: post-flare coronal rain}

\author[0000-0001-5045-827X]{Wenzhi Ruan}
\affiliation{Centre for mathematical Plasma Astrophysics, Department of Mathematics, KU Leuven, \\
Celestijnenlaan 200B, B-3001 Leuven, Belgium}

\author{Yuhao Zhou}
\affiliation{Centre for mathematical Plasma Astrophysics, Department of Mathematics, KU Leuven, \\
Celestijnenlaan 200B, B-3001 Leuven, Belgium}

\author[0000-0003-3544-2733]{Rony Keppens}
\affiliation{Centre for mathematical Plasma Astrophysics, Department of Mathematics, KU Leuven, \\
Celestijnenlaan 200B, B-3001 Leuven, Belgium}

\begin{abstract}

Most solar flares demonstrate a prolonged, hourlong post-flare (or gradual) phase, characterized by arcade-like, post-flare loops (PFLs) visible in many extreme ultraviolet (EUV) passbands. These coronal loops are filled with hot -- $\sim 30 \,\mathrm{MK}$ -- and dense plasma, evaporated from the chromosphere during the impulsive phase of the flare, and they very gradually recover to normal coronal density and temperature conditions. During this gradual cooling down to $\sim 1 \,\mathrm{MK}$ regimes, much cooler -- $\sim 0.01 \,\mathrm{MK}$ -- and denser coronal rain is frequently observed inside PFLs. Understanding PFL dynamics in this long-duration, gradual phase is crucial to the entire corona-chromosphere mass and energy cycle. Here we report a simulation in which a solar flare evolves from pre-flare, over impulsive phase all the way into its gradual phase, which successfully reproduces  post-flare coronal rain. This rain results from catastrophic cooling caused by thermal instability, and we analyse the entire mass and energy budget evolution driving this sudden condensation phenomenon. We find that the runaway cooling and rain formation also induces the appearance of dark post-flare loop systems, as observed in EUV channels. We confirm and augment earlier observational findings, suggesting that thermal conduction and radiative losses alternately dominate the cooling of PFLs.

\end{abstract}


\section{Introduction}

Solar flares represent explosive phenomena in the solar atmosphere, where $10^{28} - 10^{32}$ ergs of energy originally stored in the solar magnetic field can suddenly be released via magnetic reconnection \citep{Sweet1958,Shibata2011}. 
The time development of a solar flare event can be divided into three phases: a preflare phase, a sudden, impulsive phase and a gradual or post-flare phase \citep{Kane1974}. 
The magnetic energy is released rapidly in the impulsive phase within a typical timescale of (tens of) minutes. A large fraction of this released energy is transported from the tenuous and hot corona downwards to the denser and colder solar chromosphere via thermal conduction and by energetic electrons \citep{Antiochos1978,Ruan2020}. This deposition of energy in the chromosphere leads to a sudden heating of the local plasma, and causes upward evaporation of the plasma to form super hot ($\sim$10 MK) and dense ($\sim 10^{10}$ cm$^{-3}$) arcade-like loop systems at coronal heights. The loops return to their usual coronal conditions ($\sim$1 MK, $\sim 10^{8} - 10^{9}$ cm$^{-3}$) in the following, gradual phase, where field-guided thermal conduction and radiative losses generally contribute to the cooling process \citep{Cargill1995,Aschwanden2001}. These loops, visible in extreme ultraviolet (EUV), but also in H$\alpha$ images, are usually called post-flare loops (PFLs) \citep{Bruzek1964}. 

Thanks to dramatically increased spatio-temporal resolutions in observations, this gradual phase of solar flares is now known to show PFLs which spontaneously develop fine-scale coronal rain \citep{Scullion2014,Martinez2014,Jing2016,Scullion2016}. 
In the multi-thermal coronal rain events, cool and dense rain blobs form in-situ in the hot corona, to fall to the chromosphere with speeds up to 100 km s$^{-1}$. 
Coronal rain in flaring coronal loops has been reproduced in 3D simulations by \citet{Cheung2019} and \citet{Chen2021}, although the dynamics of this rain has not been analyzed in any detail.
Coronal rain is also observed in non-flaring coronal loops, is frequently found in loops of active regions \citep{Leroy1972,Levine1977,Schrijver2001,O'Shea2007,Antolin2012,Ahn2014,Antolin2015}, and this type of coronal rain has been studied previously using magnetohydrodynamic (MHD) simulations \citep{Fang2013,Fang2015a,Moschou2015,Xia2017,Kohutova2020}. 
\citet{Kohutova2020} successfully reproduces coronal rain in a self-consistent 3D radiative MHD simulation, in which the radiative loss in chromosphere and corona is offset by Ohmic and viscous heating.
It has been generally accepted that these rain blobs are generated in a catastrophic cooling process, essentially caused by thermal instability \citep{Parker1953,Field1965,Claes2019,Claes2020}. In a catastrophic cooling event, local temperatures drop from 1 MK to below 0.1 MK within one minute, while local densities can increase by orders of magnitude \citep{Scullion2016}. Observations of flare-driven coronal rain demonstrate that this catastrophic cooling can also happen in PFLs, but this has thus far never been modeled in detail.
 Another phenomenon thought to have a close relationship with these sudden condensations are the so-called dark post-flare loops (DPFLs), where some PFL loops suddenly vanish from specific EUV passbands, e.g. at 17.1 nm, 30.4 nm and 21.1 nm \citep{Song2016,Jejcic2018,Heinzel2020}. In these DPFLs, an EUV loop which was bright for a while suddenly darkens for several minutes, so effectively disappears between the adjacent EUV loops seen at the same height. It has been suggested that the formation of cool and dense coronal rain may contribute to EUV emission and absorption, inducing DPFL formation \citep{Jejcic2018,Heinzel2020}. However, this suggestion must still be confirmed in an ab-initio model.

Here we perform an MHD simulation of a flare event from its pre-flare phase all the way into the gradual phase. This simulation finally allows us to understand the complex thermodynamic evolutions of PFLs. 
We (1) reproduce post-flare coronal rain; (2) quantify the chromosphere-corona mass and energy cycles during PFLs;  and (3) demonstrate the intricate relationship between condensations and the disappearing EUV loops, or DPFLs.

\section{Methods} 

We perform the simulation with the open-source MPI-AMRVAC code \citep{Xia2018,Keppens2021}. The simulation is 2.5D, where the domain is 2D but all vector quantities have three components. The simulation domain is given by -75 Mm $\leq x \leq$ 75 Mm and  0 $\leq y \leq$ 100 Mm. This simulation box has an initial resolution of $96 \times 64$, but an equivalent high resolution of 6144 $\times$ 4096 is achieved with our block-adaptive mesh. The governing equations are the magnetohydrodynamic (MHD) equations with effects of gravity, thermal conduction, radiative loss and magnetic field dissipation due to resistivity included, also shown in \citet{Ruan2020} (the source terms related to fast electrons have not been activated here). The new multi-dimensional field-line-based transition region adaptive conduction (TRAC-L) method is adopted to properly handle the chromosphere-corona interaction at affordable resolution \citep{Zhou2021}.

A relaxation has been done to obtain a static pre-flare atmosphere before we perform the flare simulation. A background heating is required to offset the energy losses due to radiative cooling and thermal conduction. Inspired by \citet{Ye2020}, this background heating is a function of initial and spatio-temporally evolving values, given by
\begin{equation}
H_b (x,y,t) = 0.5 \{\tanh[(y-h_{tra})/h_a] + 1\} N_{\mathrm{e},0} (y) N_{\mathrm{e}} (x,y,t) \left( \frac{T_0 (y)}{T (x,y,t)} \right)^2 \Lambda(T_0 (y)),
\label{bQ}
\end{equation}
where $h_{tra}=3$ Mm, $h_a=0.1$ Mm, $N_{\mathrm{e}}$ indicates electron number density, subscript 0 indicates the $t=0$ initial value and $\Lambda(T)$ is the radiative cooling curve adopted in our simulations. The cooling curve from \citet{Colgan2008} is used. 
 Many other optically thin cooling curves, such as from \citet{Schure2009}, are also available in MPI-AMRVAC. \citet{Hermans2021} showed that changing the cooling curve does not truly change whether condensations happen, but only impacts the growthrate of the thermal instability and the morphology of the formed condensations.
The initial vertical temperature $T_0(y)$ profile in \citet{Avrett2008} (model C7) is employed. The number density at $y=40$ Mm is set to $2\times10^9$ cm$^{-3}$ and the initial density profile is calculated based on hydrostatic equilibrium. In the relaxation stage, a uniform vertical magnetic field is adopted. A numerically static atmosphere is obtained after a relaxation time corresponding to 3.5 hours. The local number density at $y=40$ Mm decreases to about $10^9$ cm$^{-3}$ after this relaxation. The final instantaneous background heating rate of this relaxed stage is saved and then used in the subsequent flare simulation.
Such a background heating ($\sim 10^{-4}$ to $\sim 10^{-3}$ erg cm$^{-3}$ s$^{-1}$) is crucial for keeping the temperature profile outside the flare region and the PFLs, but it is not impacting strongly the condensation process. We verified that condensations can still happen without this added background heating, with an additional simulation, which can be found on the website of ERC PROMINENT project (\url{https://erc-prominent.github.io//media/Ruan/}). 

After the relaxation, the magnetic field configuration is changed. The magnetic configuration from \citet{Ye2020} is employed then, given by
\begin{eqnarray}
B_x &=& 0, \\
B_y &=&
\begin{cases}
-B_0, & \quad x<-\lambda \\
B_0, & \quad x>\lambda \\
B_0 \sin [\pi x/(2\lambda)], & \quad {\mathrm{else}} \\
\end{cases}
\\
B_z &=& \sqrt{B_0^2-B_y^2},
\end{eqnarray}
where $B_0=30$ G is the new initial magnetic field strength and $\lambda=10$ Mm. Such a configuration allows magnetic reconnection. A three-stage resistivity strategy is then activated. A spatially localized resistivity inside the initial current sheet triggers magnetic reconnection in the first stage. This localized resistivity is given by
\begin{equation}
\eta (x,y,t<t_{\eta 1}) = 
\begin{cases}
\eta_1 [2 (r/r_{\eta})^3 - 3 (r/r_{\eta})^2 + 1] , & \quad r \leq r_{\eta} \\
0 , & \quad r>r_{\eta}
\end{cases}
\end{equation}
where $\eta_1=0.1$, $r=\sqrt{x^2+(y-h_{\eta})^2}$, $h_{\eta}=40 \ \rm Mm$, $r_{\eta}=2.4 \ \rm Mm$ and $t_{\eta 1}=31 \ \rm s$. In the second stage, we use an anomalous resistivity given by
\begin{equation}
\eta (x,y, t_{\eta 1} < t < t_{\eta 2}) = 
\begin{cases}
0 , & \quad v_d \leq v_c \\
\textrm{min}\{ \alpha_{\eta} (v_d/v_c-1) \exp[(y-h_{\eta})^2/h_s^2] , 1\}, & \quad v_d > v_c
\end{cases}
\end{equation}
where $\alpha_{\eta}=1\times 10^{-3}$, $h_s=10 \ \rm Mm$, $t_{\eta 2}=7.78$ min, $v_d(x,y,t)=J/(e N_{\mathrm{e}})$ and $v_c=128,000 \ \rm km\ s^{-1}$. The resistivity is set to zero in the third stage $t>t_{\eta 2}$ to force the flare to enter the gradual phase, where only numerical dissipation happens. This means that the explicit influence of Joule heating on the cooling of PFLs and the triggering of the condensation is neglected. The resistivity strategy used in our first and second stage is similar to that in \citet{Yokoyama2001}.

We employ symmetric boundary conditions for number density,  pressure and magnetic field components, while anti-symmetric conditions are employed for velocity components at the left and right boundaries. At the upper and bottom boundaries, density and pressure are fixed to their initial values. The magnetic field components at the bottom boundary are also fixed to the initial values.  An anti-symmetric condition is applied for the $x$-component of the magnetic field at our upper boundary, while the other two components of the magnetic field employ symmetric conditions there. The velocity at the upper boundary is set to zero, as we are not interested in an erupting fluxrope structure here, but focus on the PFLs. Anti-symmetric conditions are employed for the velocity components at the bottom boundary.

The SXR emission is calculated with the method reported in \citet{Pinto2015}. The EUV emissions are calculated with the contribution function provided by the CHIANTI database and the optically thin assumption \citep{DelZanna2015}.

\section{Post-flare loop formation and evolution} 

\begin{figure}[ht!]
\plotone{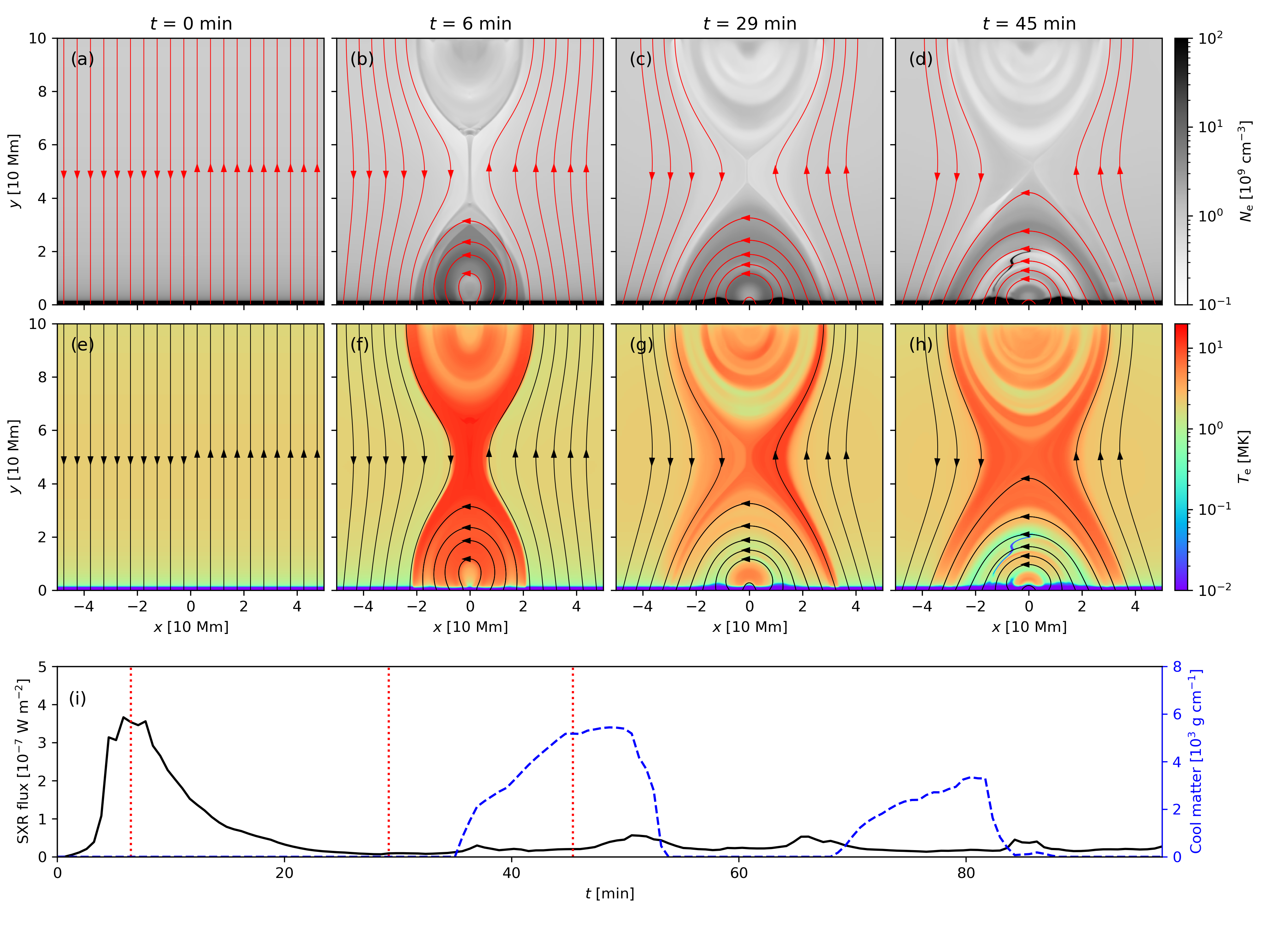}
\caption{(a-d): Electron number density (background color map) and magnetic field topology (in red) at $t=$ 0, 6, 29 and 45 min. (e-h): Evolution of temperature. (i): Temporal evolution of the integral SXR flux (black solid line) and of the total coronal rain matter (blue dashed line). The PFLs have an assumed width of 100 Mm in the invariable $z$-direction, to calculate the SXR flux. The times corresponding to the top panels are indicated in panel (i) with vertical dotted red lines, only the last panels (d-h) show postflare coronal rain. An animation of this figure is available, showing the evolution of electron number density and magnetic field configuration. It covers 97.25 minutes starting at $t=0$. The realtime duration of the animation is 15 s. \label{PFLs_evl}}
\end{figure}

Fig.~\ref{PFLs_evl} demonstrates the full evolution in magnetic topology, from pre-flare current sheet to the formation of PFLs at the impulsive phase (Fig.~\ref{PFLs_evl}b), extended to the entire evolution of the PFLs through the gradual phase (Fig.~\ref{PFLs_evl}c,d). There is a vertical current sheet separating regions of opposite field directions at the beginning of our simulation (Fig.~\ref{PFLs_evl}a). Magnetic reconnection inside this current sheet produces closed magnetic arcades below and a flux rope above a reconnection site in the impulsive phase ($t\lesssim 8$ min). 
The magnetic energy is converted into heat at the reconnection site and then conducted into the chromosphere. Deposition of the thermal energy there leads to a sudden increase of local pressure and this produces upward evaporation flows.
They fill the generated coronal loops with hot plasma ($\sim$10 MK) and increase the electron number density by one order of magnitude in the PFLs (Fig.~\ref{PFLs_evl}b,f). Thereafter, the flare enters the gradual phase when we have a rapidly decreasing magnetic reconnection speed ($t\gtrsim 8$ min). The PFL temperature decreases slowly due to thermal conduction and radiative losses in this gradual phase, but suddenly triggers thermal instability near $t \simeq 35$ min. The loop density also decreases in this period, but is still much higher than the external coronal density (Fig.~\ref{PFLs_evl}c). 
Energetic electron beam deposition can also lead to chromospheric evaporations, but this mechanism is less efficient in weak flares where the beam energy flux is small \citep{Fisher1985}. Electron beam heating has not been included into our simulation for this reason, though this heating has been successfully reproduced in a 2.5D flare simulation in our previous work \citep{Ruan2020}. The peak flux in our simulation is about $4\times 10^{-7}$ W m$^{2}$ when assuming that the loop depth in the third direction is 100 Mm (see Fig. \ref{PFLs_evl}i), therefore the simulated flare is a B level flare.

In the impulsive phase, the energy gain of the flare loop is determined by the reconnection process whereas the energy loss is mainly handled by thermal conduction, which overwhelms the radiative cooling by more than one order. Therefore, the flare temperature can be predicted from the coronal magnetic field strength, density and flare loop length or simply from coronal plasma $\beta$ with the scaling laws provided in \citet{Yokoyama1998,Yokoyama2001} and \citet{Takasao2015}. This flaring temperature estimated from equation~(1) of \citet{Yokoyama1998} well matches our looptop temperature in the impulsive phase ($\sim$20 MK), where a combination of coronal magnetic field strength 30 G, coronal number density 10$^9$ cm$^{-3}$ and half loop length 25 Mm gives an estimated temperature of 26 MK.



Catastrophic cooling driven by thermal instability condenses local plasma in-situ and leads to high density (close to 10$^{11}$ cm$^{-3}$) and cold (close to 0.01 MK) structures in the coronal PFLs (Fig.~\ref{PFLs_evl}d,h). Fig.~\ref{PFLs_evl}i shows how the SXR flux reaches its peak value at the impulsive phase, to then gradually decrease as the loop temperature drops. The temporal evolution of the total coronal rain material (i.e. $T_{\mathrm{e}} < 0.1$ MK, $N_{\mathrm{e}} > 10^{10}$ cm$^{-3}$ and $y > 5$ Mm) is also illustrated in Fig.~\ref{PFLs_evl}e. We note that sudden condensations happen in two successive events during the entire simulated period. These are located in different loop systems, with the second rain event appearing at a higher altitude.

\begin{figure}[ht!]
\plotone{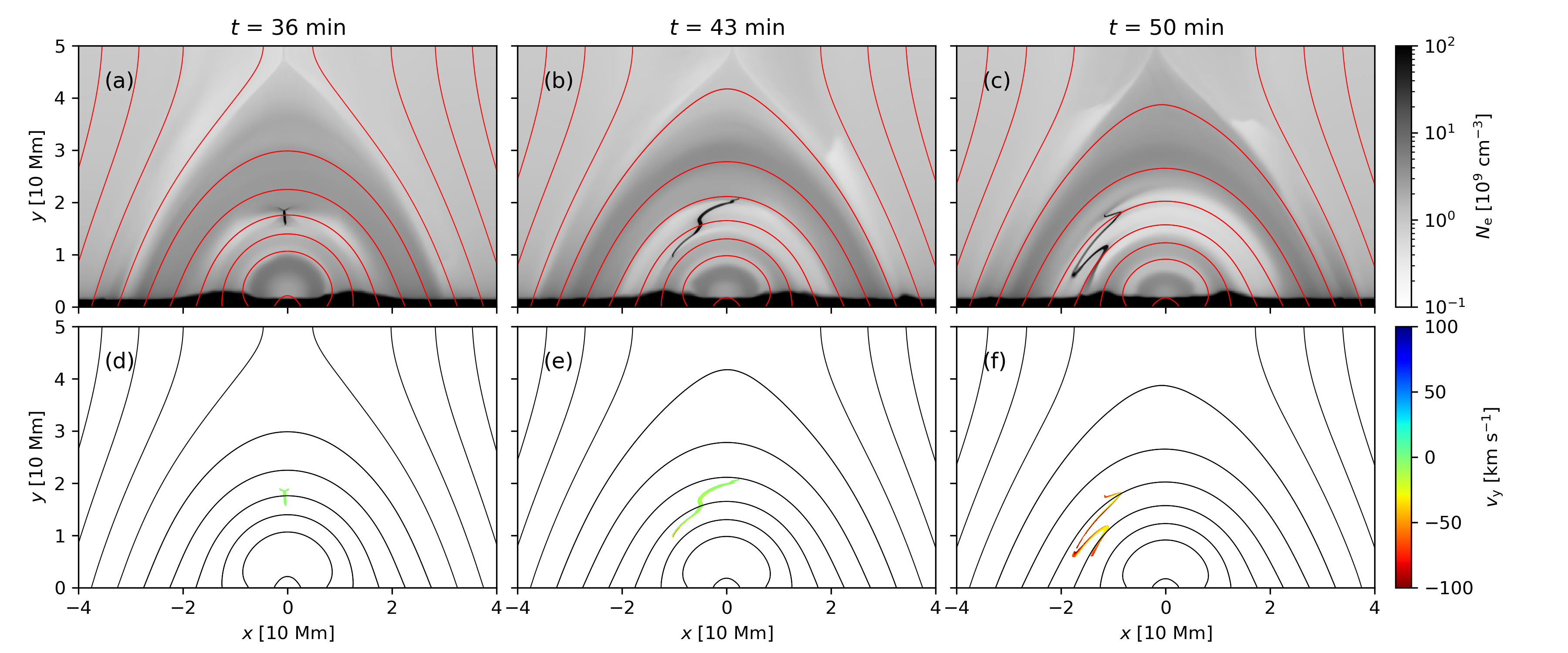}
\caption{(a-c): Electron number density at $t=$ 36, 43 and 50 min. (d-f): Vertical $y$-component of velocity for the coronal cool plasma ($T_{\mathrm{e}} < 0.1$ MK, $N_{\mathrm{e}} > 10^{10}$ cm$^{-3}$ and $y > 5$ Mm). The solid lines are magnetic field lines. An animation of this figure is available, showing the evolution of electron number density and $y$-component of velocity for the cool plasma. It covers 32.41 minutes starting at $t=32.42 \ \textrm{min}$. The realtime duration of the animation is 5 s. \label{Coronal_rain}}
\end{figure}

Once condensations happen within PFLs, the formed cold and dense plasma structures will likely fall down from coronal heights due to gravity and hence appear as observed coronal rain blobs. This is fully reproduced in our simulation as demonstrated in Fig.~\ref{Coronal_rain}. Cold plasma is formed at a PFL looptop at the beginning of the runaway condensation (Fig.~\ref{Coronal_rain}a). 
Thereafter, the cold structure extends to lower and higher loops, meanwhile sliding down to one side along the magnetic post-flare arcade (Fig.~\ref{Coronal_rain}b,c). This falling cold plasma gets accelerated to a speed of $\sim$100 km s$^{-1}$ by gravity before it enters the chromosphere (Fig.~\ref{Coronal_rain}d-f). Such a speed is close to that found in coronal rain observations \citep{Martinez2014}. Considering an acceleration timescale of 10 minutes, the average acceleration rate is lower than the acceleration of gravity. A detailed analysis of rain blob acceleration process for non-flaring (or quiescent) coronal rain was given in \citet{Fang2013}.
Different from \citet{Fang2013}, the guide field B$_z$ of the loops where condensation happens is nearly zero. But since coronal rain is formed through thermal instability which depends only slightly on the magnetic field, whether or not there is a guide field should have little influence on our results.

\section{Mass and energy cycles during the gradual phase}

\begin{figure}[ht!]
\plotone{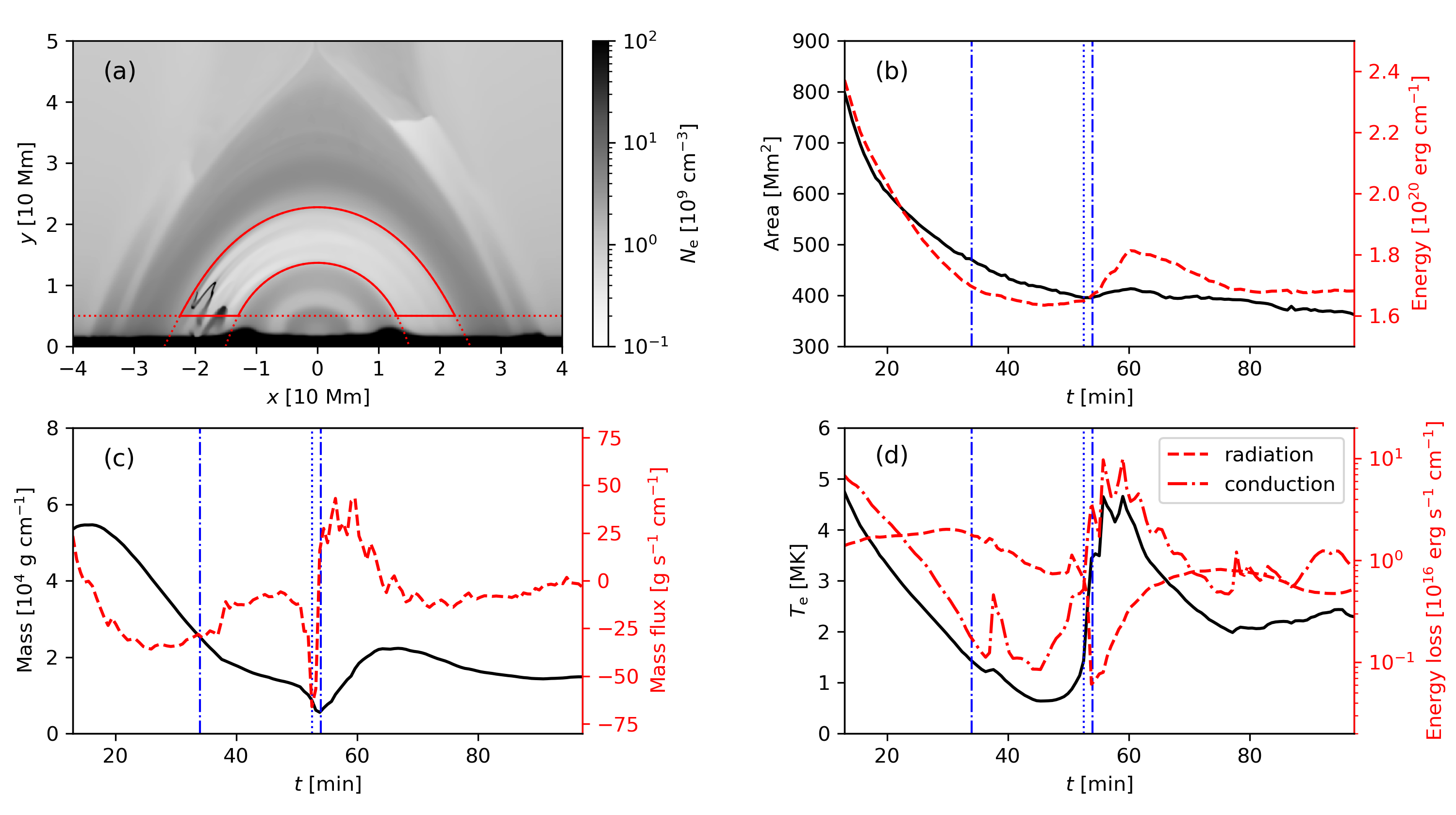}
\caption{(a): Electron number density at $t =$ 52 min. The region bounded by the field lines starting from $(x,y)=(-25\ \mathrm{Mm},0)$ and $(x,y)=(-15\ \mathrm{Mm},0)$ and the horizontal line $y=$5 Mm is investigated in panels b-d. (b): Time evolution of the evolving area (black solid line) and that of integrated total energy (red dashed line). (c): Time evolution of integrated mass (black solid line) and of the mass flux across the lower boundaries of this region. (d): Time evolution of average temperature (black solid line), integral radiative losses (red dashed line) and integral conductive losses (red dashed-dotted line). Blue vertical dashed-dotted lines indicate the starting and ending time of the condensation, and the vertical blue dotted line marks the time of panel a. An animation of this figure is available, showing the evolution of electron number density and boundaries of the investigated region. It covers 84.28 minutes starting at $t=12.97 \ \textrm{min}$. The realtime duration of the animation is 13 s. \label{Mass_cycle}}
\end{figure}

Here we investigate the mass and energy cycles in our entire 100-minute simulation of the gradual phase and the role of condensations in it. To do so, we track mass and energy budgets into the coronal part ($y >$ 5 Mm) of a loop section in which the first round of condensation happens. This loop section is always bounded by (a) the evolving magnetic field line with a fixed footpoint at $x=-25\ \mathrm{Mm}$ at our lower $y=0$ boundary; and by (b) a similarly evolving field line with footpoint at $x=-15\ \mathrm{Mm}$ (Fig.~\ref{Mass_cycle}a). As seen in Fig.~\ref{Coronal_rain} and Fig.~\ref{Mass_cycle}a, this region gets emptied during the first round of condensations, as matter collects along the field lines into localized rain blobs. This entire loop system continuously moves downward as a result of the above reconnection dynamics and the corresponding area of the selected region continuously decreases during the simulation as quantified by the solid line in Fig.~\ref{Mass_cycle}b. The area-integrated total energy (kinetic, thermal and magnetic combined, dashed line in Fig.~\ref{Mass_cycle}b), area-integrated mass (solid line in Fig.~\ref{Mass_cycle}c) and the area-averaged temperature (solid line in Fig.~\ref{Mass_cycle}d) of this region also continuously decrease before the first condensation ($t$ $\simeq$ 35 min). 
Plasma which evaporated upwards into the corona in the previous impulsive phase now leaks back to the chromosphere in this period, seen in the downward mass flux (dashed line) in Fig.~\ref{Mass_cycle}c. The decrease of temperature leads to a decrease of the atmospheric scale height. However, this downwards leakage of coronal plasma is severely reduced when the condensation happens after $t\gtrsim$ 35 min, since the gas pressure in the coronal part of the loop then decreases rapidly. Therefore, we see a drop of the downward mass flux in Fig.~\ref{Mass_cycle}c near $t$ $\approx$ 38 min, when condensations fully formed. 
Later on, the downward mass flux experiences a sudden increase, exactly when the cool coronal rain material goes through the lower $y=5$ Mm boundary of the studied region. 
The rain material is dense and heavy, so the material can still drop even if there is an upward pressure gradient below it.
The area-integrated mass reaches its minimum value when all cool material leaves and enters the chromosphere. Thereafter, plasma from the chromosphere is injected to the loop again, due to the low pressure inside the loop, leading to an upwards mass flux. The total mass then gradually returns to its value before condensation. 

The changing coronal energy budget shows a similar tendency with the changing coronal mass cycle. The energy also experiences a decrease in the pre-condensation and during the condensation phase, to then experience an increase after the condensations merged into the chromosphere due to renewed plasma injection (Fig.~\ref{Mass_cycle}b). It has been suggested that thermal conduction determines the PFLs energy loss at the beginning of the gradual phase of a flare and that subsequently radiative losses will become dominant \citep{Cargill2004}. Our simulation shows that this suggestion is correct before and also during the occurring condensations. The efficiencies of radiative cooling and of thermal conduction to the energy loss in the selected region are compared in Fig.~\ref{Mass_cycle}d. The contribution of thermal conduction is greater than radiative losses for $t \lesssim 25$ min, but conductive losses drop gradually owing to the decreasing temperature gradient. 
At $t \approx$ 25 min, the average temperature is about 3 MK, and the efficiency of radiative losses becomes most prominent. 
However, conductive losses become stronger than radiative losses again when the condensations vanished from the loop system, as collisions between re-injected flows from both footpoints make the loop hot again and the radiative losses drop for a while due to the decrease of loop density.

\section{Catastrophic cooling and rain-induced QPP}

\begin{figure}[ht!]
\plotone{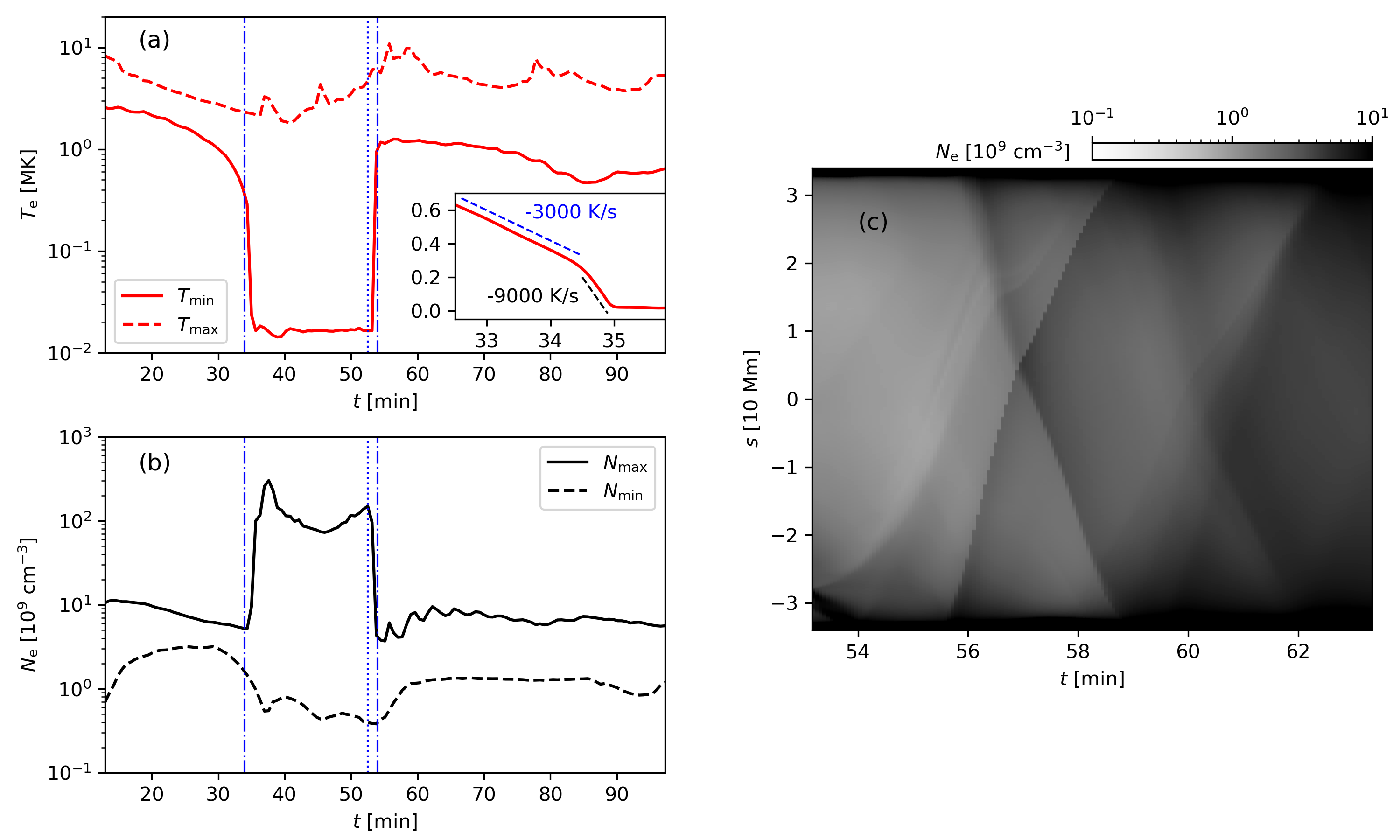}
\caption{(a): Time evolution of maximum/minimum temperature in the region showing coronal rain from Fig.~\ref{Mass_cycle}a. (b): Evolution of maximum/minimum number density in the same region. (c): Time-space plot of the number density along a field line with $y=0$ footpoints at $x=\pm$ 24.5 Mm, after the rain left the studied loop region, and a quasi-periodic oscillation appears. The midpoint of the field lines is at $s=0$ and negative $s$ indicates the left side.  \label{condensation}}
\end{figure}

The first round of condensation happens near $t \approx 35$ min. The temporal evolutions of instantaneous maximum/minimum temperature/number density in the condensation region (the same as marked in Fig.~\ref{Mass_cycle}a) are illustrated in Fig.~\ref{condensation}. 
Triggering of thermal instability switches the radiative cooling process from linear to nonlinear, and then leads to a catastrophic cooling of local plasma. We get an average temperature decreasing rate of -9000 K s$^{-1}$ in the catastrophic cooling phase. In contrast, the cooling rate before catastrophic cooling is -3000 K s$^{-1}$. As a result of catastrophic cooling, the local temperature decreases from 0.2 MK to 0.02 MK within half a minute (Fig.~\ref{condensation}a), while the local number density increases by one order, from 10$^{10}$ cm$^{-3}$ to 10$^{11}$ cm$^{-3}$ (Fig.~\ref{condensation}b). 
Noting that the cooling rate is density dependent, the cooling rate we get in the catastrophic cooling phase is in the same order as that observed in \citet{Scullion2016} ($-22700\ \mathrm{K\ s^{-1}}$).

A quasi-periodic pulsation (QPP) with a period of $\sim$ 3 minutes appears in the maximum density curve, just after the rain condensation disappeared from the PFL system. This QPP is caused by the injected flows mentioned in the previous section, refilling and reheating the PFL. The density variation due to these flows along a field line is shown in Fig.~\ref{condensation}c. Injected flows propagating from one footpoint to the other produce reflected slow mode waves. Such a process has previously been studied in \citet{Fang2015b} for isolated loop systems. The sharp density changes in Fig.~\ref{condensation}c are shocks ahead of the injection flows and the wave fronts of the slow mode waves. Such QPPs hence reflect density variations in the low corona due to combined flows and wave propagation. The period of our QPP is close to the time for the slow mode wave to propagate from one footpoint to the other, as the wave speed is about 300 km s$^{-1}$.

\section{Coronal rain and dark post-flare loops}

\begin{figure}[ht!]
\plotone{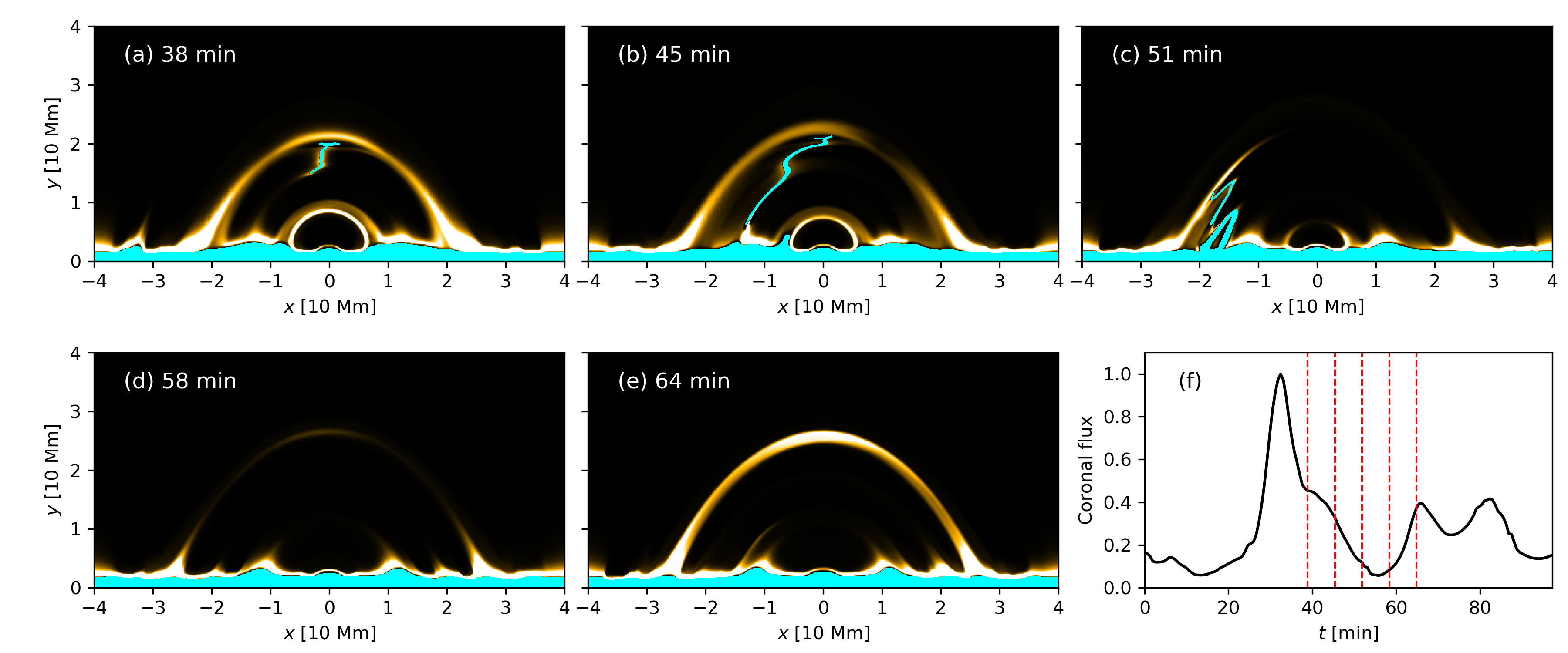}
\caption{(a-e): Time evolution of synthetic EUV 17.1 nm images. The regions in cyan have temperatures lower than 0.1 MK. (f): Time evolution of the integral EUV 17.1 nm flux from a region $y > 5$ Mm. Red vertical dashed lines in panel (f) give the corresponding times of panels (a-e). An animation of this figure is available, showing the evolution of EUV 17.1 nm emission and electron temperature. It covers 25.93 minutes starting at $t=38.90 \ \textrm{min}$. The realtime duration of the animation is 4 s.  \label{DPFLs}}
\end{figure}

In synthesized EUV images of our simulation, coronal loop(s) appear in the gradual phase. An example is shown in  Fig.~\ref{DPFLs} at the 17.1 nm waveband. Interestingly, this loop disappears for about 10 minutes during the evolution, as demonstrated in panels (a) to (e) of Fig.~\ref{DPFLs}. The sudden darkening of the bright EUV loop in our simulation resembles the dark post-flare loop (DPFL) phenomenon, previously observed at the same 17.1 nm passband and with similar timescale of 10 minutes \citep{Song2016}. Observed DPFLs and the disappearing EUV coronal loop in our simulation also share the same time evolution of integral EUV flux: the EUV flux reaches its minimum value when the darkening happens (compare Fig.~3 in ref.~\citet{Song2016} and our Fig.~\ref{DPFLs}f). {The formation of a darkened coronal EUV loop needs to satisfy one or both of the following conditions: (1) an emission drop in an existing bright EUV loop; (2) an absorption of the background EUV emission \citep{Anzer2005}.} Here we explain how these conditions can be satisfied based on our simulation results.

The role of cool and dense coronal plasma in EUV emission and absorption leading to DPFLs has been emphasized previously \citep{Jejcic2018,Heinzel2020}, with coronal rain observations \citep{Scullion2014,Martinez2014,Jing2016,Scullion2016} and our simulation results showing how this cool and dense plasma can be generated in PFLs. The drop in the loop emission is understood from our simulation: loop temperature changes relate to nearby rain condensations. Indeed, the temperature of bright loops in this 17.1 passband (about 10$^{5.8}$ K) is not far from the critical temperature for the onset of catastrophic cooling (this is density and temperature dependent, but generally happens below 2 MK according to \citet{Cargill2013}). Condensations can be triggered by thermal instability near bright loops and these suddenly formed structures grow fastest across magnetic field lines (counterintuitive due to the field-aligned thermal conduction, but see \citet{Fang2013,Claes2019,Claes2020,Zhou2020}). Rain that forms near (Fig.~\ref{DPFLs}a-b), and ultimately inside (Fig.~\ref{DPFLs}c) the bright coronal loops, is thus causing the darkening as illustrated in Fig.~\ref{DPFLs}c-d. Once condensation happens, a lot of plasma will collect into a small region, so the plasma density elsewhere in the loop decreases. To maintain pressure balance, these evacuated loop regions will increase in temperature, so EUV brightness decreases due to these combined temperature and density changes. Ultimately, the loop refills and brightens once more (Fig.~\ref{DPFLs}e).

\section{Conclusion and discussion}

To fully understand coronal rain in PFLs and the mass and energy budget in the gradual phase of solar flares, we performed a flare simulation from onset all the way into the long duration post-flare phase. Post-flare coronal rain successfully and, at least in our simulation, repeatedly forms. The flare-induced rain is a result of catastrophic cooling by thermal instabilities, and our simulation shows successive rain formation at increasing heights in the PFL configuration. Falling rain blobs into the chromosphere lead to sudden mass drops in PFLs, but their mass increases again due to spontaneously forming injection flows. Therefore, the coronal rain events do not accelerate the PFL mass loss in the longer term. Such longer term mass loss is more determined by the change of the gravity scale height due to the cooling of PFLs. 

Both thermal conduction and radiative losses contribute to the energy budget in PFLs. Thermal conduction dominates the PFL energy loss at the beginning of the gradual phase. Thereafter, it becomes less efficient than radiative losses, owing to decreases in loop temperature and in temperature gradient. However, thermal conduction can efficiently recover again after a condensation falls to the chromosphere, as the loop reaches again a high temperature. In this phase, an emptied loop refills and can show a slow-wave related QPP.

We showed that the formation of DPFLs can result from post-flare rain condensations. Condensations change loop temperatures and can make existing bright EUV loops temporarily disappear for several minutes. This timescale of EUV loop darkening is identical to observed DPFLs. 

A recent paper by \citet{Reep2020} suggests that additional energy supplement (heating) is necessary in the formation of flare-driven coronal rain based on a 1D parameter study. In their simulations, a 1D flare loop is filled with hot/dense plasma by non-thermal electron-beam-driven chromospheric evaporations at the beginning and then gradually cools down due to radiative loss. The loop will directly cool down to chromospheric temperature without condensations happening if there is no extra heating provided during the cooling, in which the coronal density also drops by orders of magnitude during the cooling. It is interesting to note that condensations can still happen inside our 2D post-flare loop, even if the background heating is switched off (see the additional simulation mentioned in section 2. Compression caused by thermal pressure of the surrounding environment or by the Lorentz force may serve as additional energy supplement, as it has been shown in Fig.~\ref{Mass_cycle}b that the loop area decreases gradually during the cooling phase. When this interaction with the surrounding environment is taken into account, the coronal temperature and density are not likely to drop by orders of magnitude at the same time as shown in \citet{Reep2020}. This difference between 1D simulations and 2D simulations of post-flare loop development will be studied in future work.

\begin{acknowledgments}
This work was supported by the ERC Advanced Grant PROMINENT, an FWO project G0B4521N and a joint FWO-NSFC grant G0E9619N. This project has received funding from the European Research Council (ERC) under the European Union’s Horizon 2020 research and innovation programme (grant agreement No. 833251 PROMINENT ERC-ADG 2018). This research is further supported by Internal funds KU Leuven, through the project C14/19/089 TRACESpace. The computational resources and services used in this work were provided by the VSC (Flemish Supercomputer Center), funded by the Research Foundation Flanders (FWO) and the Flemish Government, department EWI.
\end{acknowledgments}

\bibliography{ms}{}
\bibliographystyle{aasjournal}

\end{document}